\documentclass[aps,reprint,amsmath,amssymb,groupaddress,superscriptaddress,pra]{revtex4-2}

\usepackage[utf8]{inputenc}
\usepackage{amsmath}
\usepackage{amsfonts}
\usepackage{amssymb}
\usepackage{physics}
\allowdisplaybreaks[4]
\usepackage{bm}
\usepackage{color}
\usepackage{graphicx}
\usepackage{soul}
\usepackage{CJK}
\usepackage{xcolor}
\usepackage{mathtools}
\usepackage{lipsum}
\usepackage{float}
\usepackage{comment}

\usepackage{xcolor}
\usepackage[colorlinks,citecolor=red,linkcolor=blue,anchorcolor=blue,urlcolor=blue]{hyperref}
\usepackage{chngcntr}

\begin{document}

\title{Low-cost demonstration of the Zeeman effect: From qualitative observation to quantitative experiments}

\author{Shao-Han Qin}
\affiliation{School of Physics and Astronomy, Beijing Normal University, Beijing, 100875, China}
\affiliation{The Experimental High School Attached To Beijing Normal University, Beijing, 100032, China}
\author{Yu-Han Ma}
 \email{yhma@bnu.edu.cn}
\affiliation{School of Physics and Astronomy, Beijing Normal University, Beijing, 100875, China}
\affiliation{Graduate School of China Academy of Engineering Physics, Beijing, 100193, China}

\begin{abstract}
The Zeeman effect, a fundamental quantum phenomenon, demonstrates the interaction between magnetic fields and atomic systems. While precise spectroscopic measurements of this effect have advanced significantly, there remains a lack of simple, visually accessible demonstrations for educational purposes. Here, we present a low-cost experiment that allows for direct visual observation of the Zeeman effect. Our setup involves a flame containing sodium (from table salt) placed in front of a sodium vapor lamp. When a magnetic field is applied to the flame, the shadow cast by the flame noticeably lightens, providing a clear, naked-eye demonstration of the Zeeman effect. Furthermore, we conduct two quantitative experiments using this setup, examining the effects of varying magnetic field strength and sodium concentration. This innovative approach not only enriches the experimental demonstration for teaching atomic physics at undergraduate and high school levels but also provides an open platform for students to explore the Zeeman effect through hands-on experience.
\end{abstract}

\maketitle

\section{Introduction}
Quantum mechanics predicts that external fields can alter quantum states. This fundamental concept is exemplified by phenomena such as the Zeeman effect and Stark effect, which demonstrate the splitting of atomic energy levels under magnetic and electric fields, respectively. The Zeeman effect, in particular, plays a crucial role in our understanding of atomic structure and interactions. The Zeeman effect, first observed by Pieter Zeeman in 1896 \cite{bakker1946fifty}, marked a significant milestone in physics. Using a high-precision Rowland grating, Zeeman observed the splitting of spectral lines in a strong magnetic field, providing crucial empirical evidence of the quantization of angular momentum and the intrinsic magnetic properties of particles. The Zeeman effect has since become a cornerstone in various branches of physics, such as condensed matter physics \cite{srivastava2015valley,sun2020large} and astrophysics \cite{crutcher2019review,kochukhov2021magnetic}. 

Modern measurement techniques, including high-resolution optical spectroscopy and advanced interferometry, have enabled increasingly precise observations of how magnetic fields influence atomic energy levels \cite{taylor2017zeeman}. 
Despite these advances in precision measurement, there remains a significant gap in the pedagogical approach to the Zeeman effect. Simple and accessible visualizations of this phenomenon are scarce \cite{dos1996absorption}, creating a substantial barrier in physics education. 
This disconnect between advanced research techniques and basic educational tools highlights the need for innovative approaches in physics education.

To bridge this gap, we present a straightforward and low-cost method to demonstrate the Zeeman effect, designed specifically for educational settings. Remarkably, this experiment allows the phenomenon to be visually observed with human eye, making the Zeeman effect more accessible and comprehensible to students. In our set-up, we have table salt added to flame and a sodium vapor lamp in front of the flame. Due to atomic absorption, there will be a shadow cast behind the flame. When a strong enough magnetic field is applied around the flame, one can observe the shadow becoming dimmer. Such a clear and convenient demonstration would not only enhance students' understanding of the complex interactions between matter and electromagnetic fields but also provide a crucial hands-on learning experience. 

\section{Setup and phenomenon\label{setup}}

\begin{figure}
\raggedright(a)
\includegraphics[width=1\columnwidth]{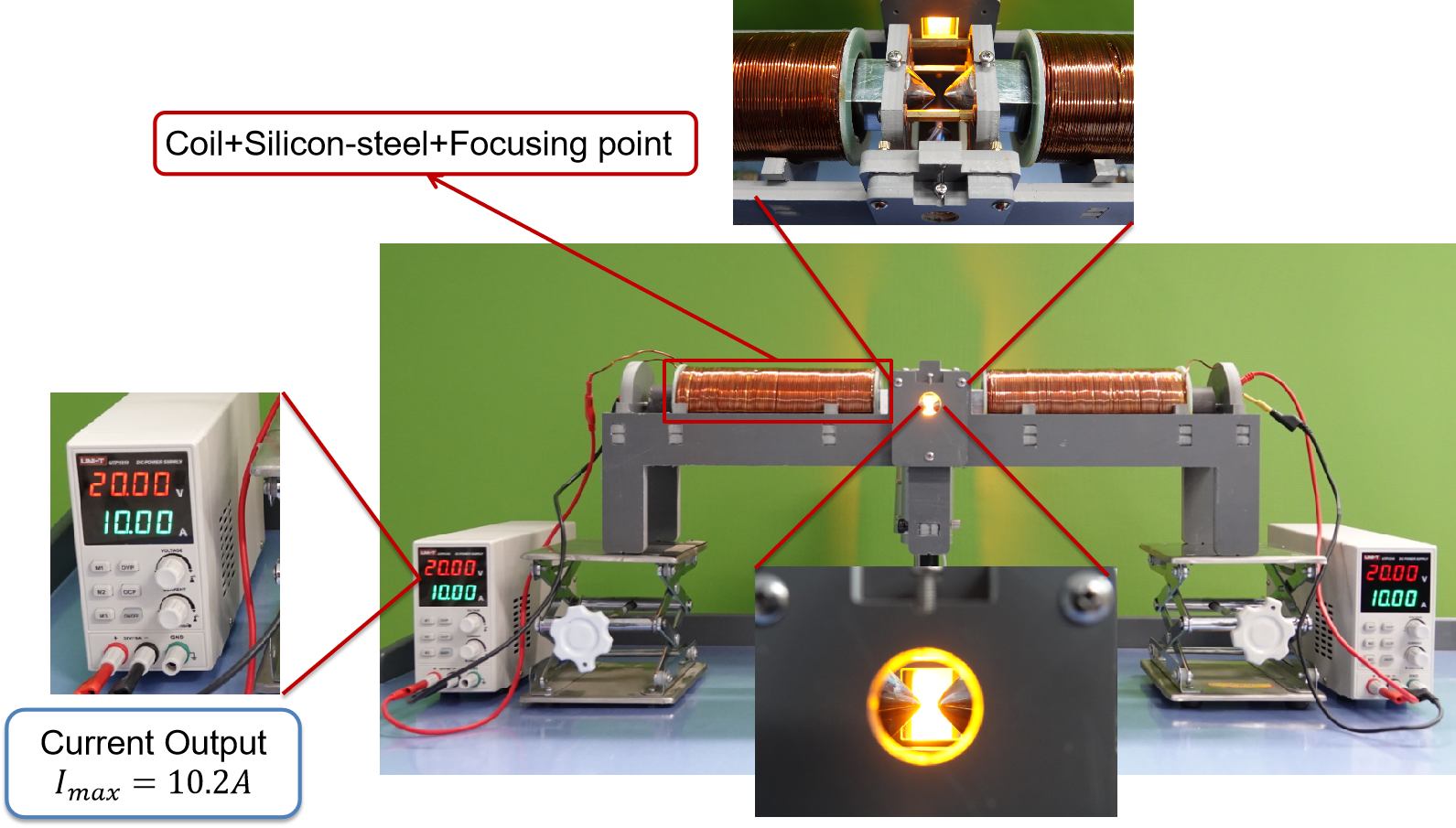}
(b)
\includegraphics[width=1\columnwidth]{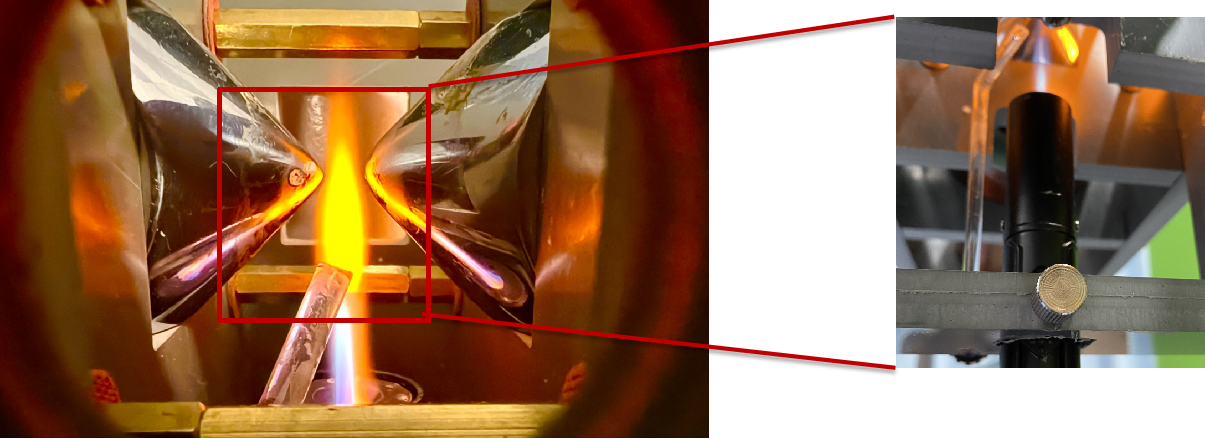}
\caption{ Electromagnet setup, with close-up views of different segments. (a) On both sides are the power supply, generating current in the electromagnet; the electromagnet is built with a silicon steel core, a focusing point at the front, and coils wound around it. Its position is fixed vertically and horizontally, and the front and top view of the electromagnet are also magnified. (b) Close-up views of the lighter from front and below show how it is located and controlled}
\label{demonstration}
\end{figure}
To significantly change the intensity of the shadow, we build an electromagnet to generate strong magnetic field in the order of 1 Tesla. As shown in Fig. \ref{demonstration}(a), the electromagnet consists of two symmetrical parts, with cores made of silicon steel, and approximately 1000 turns of copper wire wrapped around the cores in 5 layers. Current would be directed to flow through the wire, creating an initial magnetic field using the solenoid structure. The silicon steel, being ferromagnetic, has large magnetization, and can generate a much larger magnetic flux density. Two spear-like focuses made of iron is added to both ends of the electromagnet on either side and act as a focus for the magnetic field. A current output with maximum magnitude of $10.2$ A is connected to the electromagnet, producing a magnetic field of around 2 Tesla. The magnetic field produced by the electromagnet is not evenly distributed in space. Simulation in Appendix \ref{sim} shows that the magnetic field is strongest in the middle, which is also the center of the flame.

In the centre between the two electromagnets is the place for the flame, created by a lighter below. In a close-up view in Fig. \ref{demonstration}(b), the lighter is firmly located and fixed, carefully controlling the position of the flame. By adjusting the knob beside the lighter, we can turn on the lighter and adjust the magnitude of the fire. A glass stick is also fixed beside the lighter and the stick is dipped in salt water before every measurement so as to control the concentration of table salt in the flame. By putting the top of the stick directly above the mouth of the lighter, we can ensure that the fire contains table salt. To change the sodium concentration in the fire, we change the concentration of the salt water.

\begin{figure}
\raggedright(a)
\includegraphics[width=1\columnwidth]{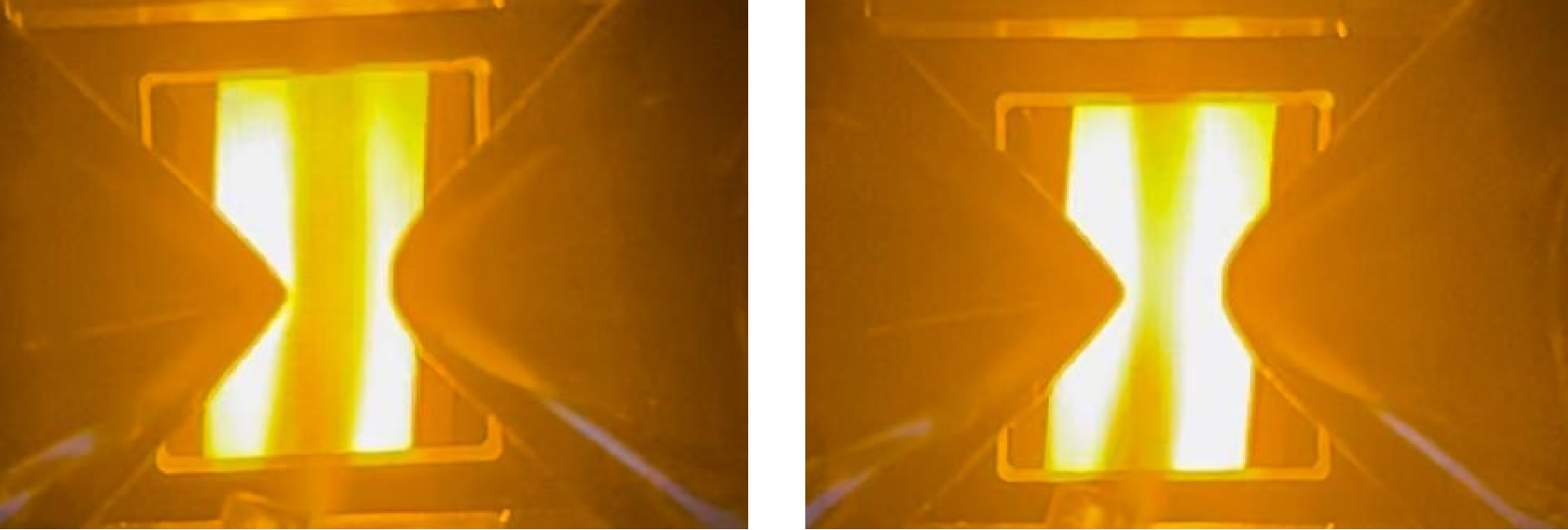}
(b)
\includegraphics[width=1\columnwidth]{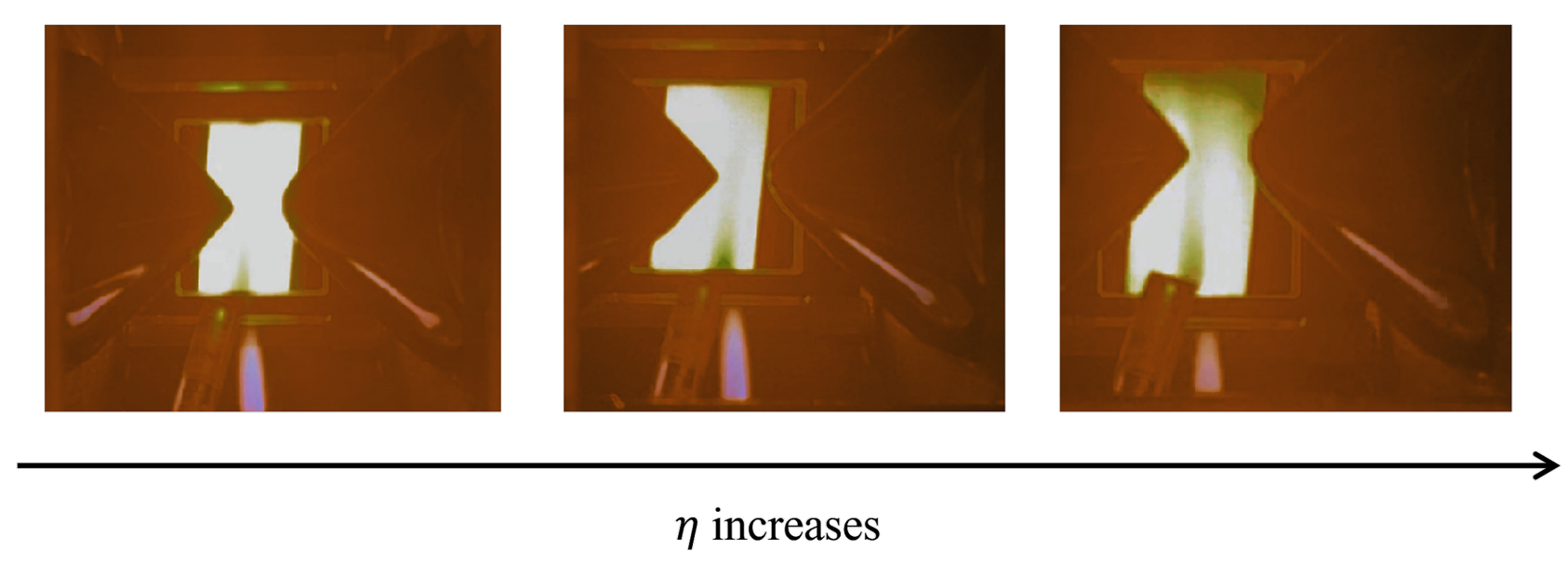}
\caption{Shadow cast by the fire with table salt. (a) Left panel: shadow without magnetic field. Right panel: shadow becomes much lighter with magnetic field being applied. (b) For a given magnetic field, shadow becomes darker as salt concentration increases from left to right, from 5\% to 15\% to 25\%}
\label{shadow}
\end{figure}

As we can see in the left plane of Fig. \ref{shadow}(a), when the magnetic field is turned off, there is an obvious shadow of the fire, which is directly captured by the camera behind the flame. Basically, the shadow is produced by the resonant absorption of sodium vapor (generated by heating table salt in the flame) in the light emitted by the sodium lamp. The right plane of Fig. \ref{shadow}(a) shows a drastic lifting of shadow after magnetic field is applied.  This is because when an external magnetic field is applied, the energy levels of electrons in the sodium atom split due to the Zeeman effect \cite{jackson1938hyperfine,foot2005atomic}. After the splitting, the absorption spectra is no longer perfectly aligned with the emission spectra of the vapor lamp, and thus less light is absorbed, creating a lighter shadow. 

We can observe that the shadow is unevenly distributed spatially, being lighter in the area between the two ends of the electromagnet than the area above or below it. This is due to the uneven magnetic field produced by the electromagnet (See Appendix \ref{sim} for details). In our quantitative measurements later on, we will measure the shadow intensity of only the area between the electromagnet's spearhead ends. This can reduce the inaccuracies caused by the shadow's unevenness. Besides, it is shown in Fig. \ref{shadow}(b) that as the concentration of salt increases, the shadow becomes darker. The reason for this is very evident. With a higher concentration and greater amount of sodium atoms in the fire, photons will more likely be absorbed, creating a darker shadow.

\section{Quantitative experiments\label{experiment}}

\begin{figure}
\includegraphics[width=1\columnwidth]{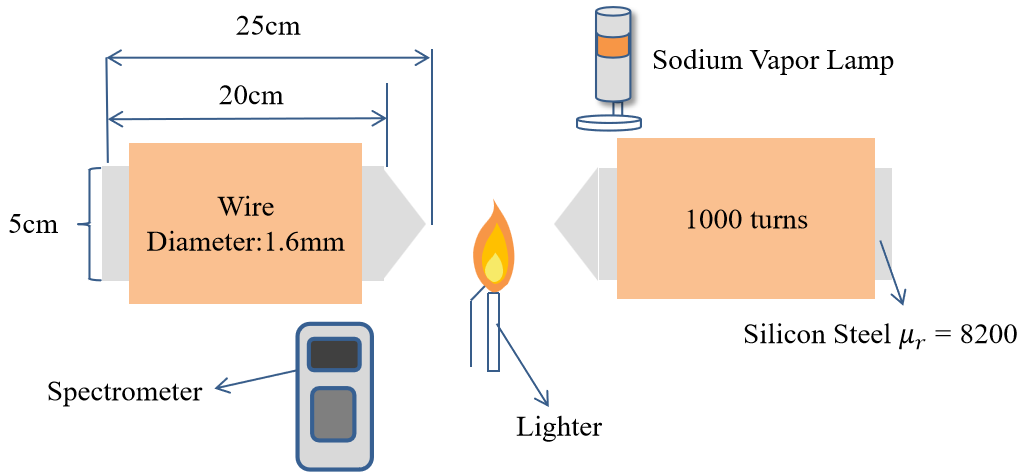}
\caption{A spectrometer is placed behind the lighter to measure the shadow intensity.}
\label{path}
\end{figure}

In addition to the demonstrative experiments with clear phenomena shown above, our setup can also be used for quantitatively investigating the Zeeman effect. As shown in Fig. \ref{path}, we setup our device to measure the shadow intensity $K_s(B,\eta)$ as a function of both magnetic field strength $B$ and salt concentration $\eta$ in detail. The light from the sodium vapor light goes through the flame of the lighter and then hits the spectrometer. The lighter produces a small but very steady flame, also narrow enough to fit into separation of the two electromagnets. Also, the dim source of light would decrease the influence of the fire’s light on the shadow brightness measurement. It is worth mentioning that we select the range of light wavelength in the spectrometer closest to 580 nm, which is the sodium’s absorption and emission spectrum. Thus, we can greatly reduce error from the background lighting and other sources of light.

\begin{figure}
\raggedright(a)
\includegraphics[width=1\columnwidth]{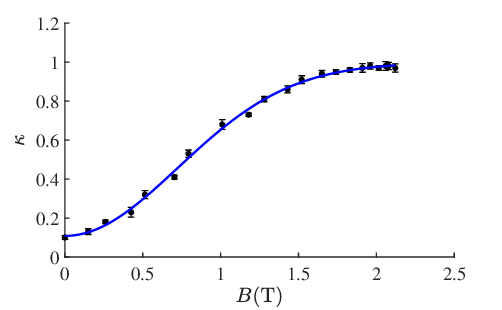}
\raggedright(b)
\includegraphics[width=1\columnwidth]{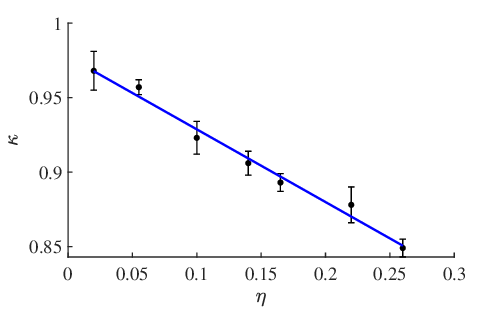}
\caption{Shadow intensity varies with the magnetic field strength (a) and the salt concentration (b). The black dots are experimental data with errorbar and the blue curve represents the fitting curve, the form of which is given by Eq. \eqref{k}.}
\label{kBC}
\end{figure}

We first measure the light intensity of the shadow as a function of magnetic field strength, which is tuned by changing the current $I$ supplied to the electromagnet. The measured $I-B$ curve is given in Appendix \ref{B-I}, and the concentration of salt water is kept at $3\%$ in this experiment. As shown in Fig. \ref{kBC}(a), we plot the shadow brightness factor $\kappa\equiv K_s(B,\eta)/K_s(0,0)$(black dots) at different magnetic field strength $B$. Here, $K_s(0,0)$ is light intensity of the shadow without adding salt and at zero magnetic field. In our experiment, each dot is the average of six independent replicates, and the corresponding error bar represent the variance of these replicates. It is seen in this figure that the brightness of the shadow increases with magnetic field, which means the shadow becomes lighter as magnetic field increases, explaining the phenomena observed. The experimental results fit well with the theoretical prediction (See Appendix \ref{k-B} for details)
\begin{equation}
\kappa(B,\eta)=1-A\eta e^{-\lambda B^{2}},
\label{k}
\end{equation}
where $A$ and $\lambda$ is the fitting parameters.

In addition, we change the concentration of salt while fixing the magnetic field at $1.5$T. For each of the concentrations measured, we use the scale to weigh the amount of salt, then add the salt into a fixed amount of water, creating an accurate mix of salt water. The salt water is then trickled onto the glass stick above the flame. It is illustrated in Fig. \ref{kBC}(b) that the shadow darkens linearly with the salt concentration. We would like to mention that the concentration of salt inside the flame isn’t precisely constant over time, thus limiting our measurement accuracy. This is reflected in the fact that the data fluctuations are more pronounced in Fig. \ref{kBC}(b) than in Fig. \ref{kBC}(a). To control the initial concentration, we soaked the glass rod into a certain mixture of salt water, but the amount of salt water on each rod is not accounted for. Improvements on controlling salt concentration and amount need further consideration.

\section{Summary\label{conclusion}}

In conclusion, with low-cost and readily available devices, we have introduced an interesting experiment demonstrating the Zeeman effect with observable phenomena visible to the human eye. Compared to conventional methods of observing spectral shifts in traditional Zeeman effect experiments, the observation of shadows in this experiment offers a more direct, concrete, and engaging advantage, particularly in educational settings where it can be more easily implemented. 

In addition to serving as a demonstration experiment for high school students or junior undergraduate students, our setup can also be used for quantitatively exploring the impact of magnetic field strength on the Zeeman effect, making it suitable as a platform for research-based learning tasks in higher-level undergraduate course or graduate teaching. The proposed experimental setup is particularly valuable in physics education, as it helps bridge the gap between abstract concepts and physical reality, fostering deeper comprehension and retention of the Zeeman effect.

\begin{acknowledgments}
    This study is inspired by the \href{https://www.iypt.org/problems/problems-iypt-2024/}{17th problem of the 2024 International Young Physicists' Tournament (IYPT)}. We are grateful to Yun-Qian Lin and Jia-Rui Lei for their helpful discussions in this work. We also appreciate the insightful comments and careful reading from Yin-Long Wang and Sai Li on this manuscript. Y. H. Ma thanks the National Natural Science Foundation of China for support under grant No. 12305037 and the Fundamental Research Funds for the Central Universities under grant No. 2023NTST017. 
\end{acknowledgments}

\appendix
\section{Simulation of electromagnet\label{sim}}
\begin{figure}[H]
\includegraphics[width=0.565\columnwidth]
    {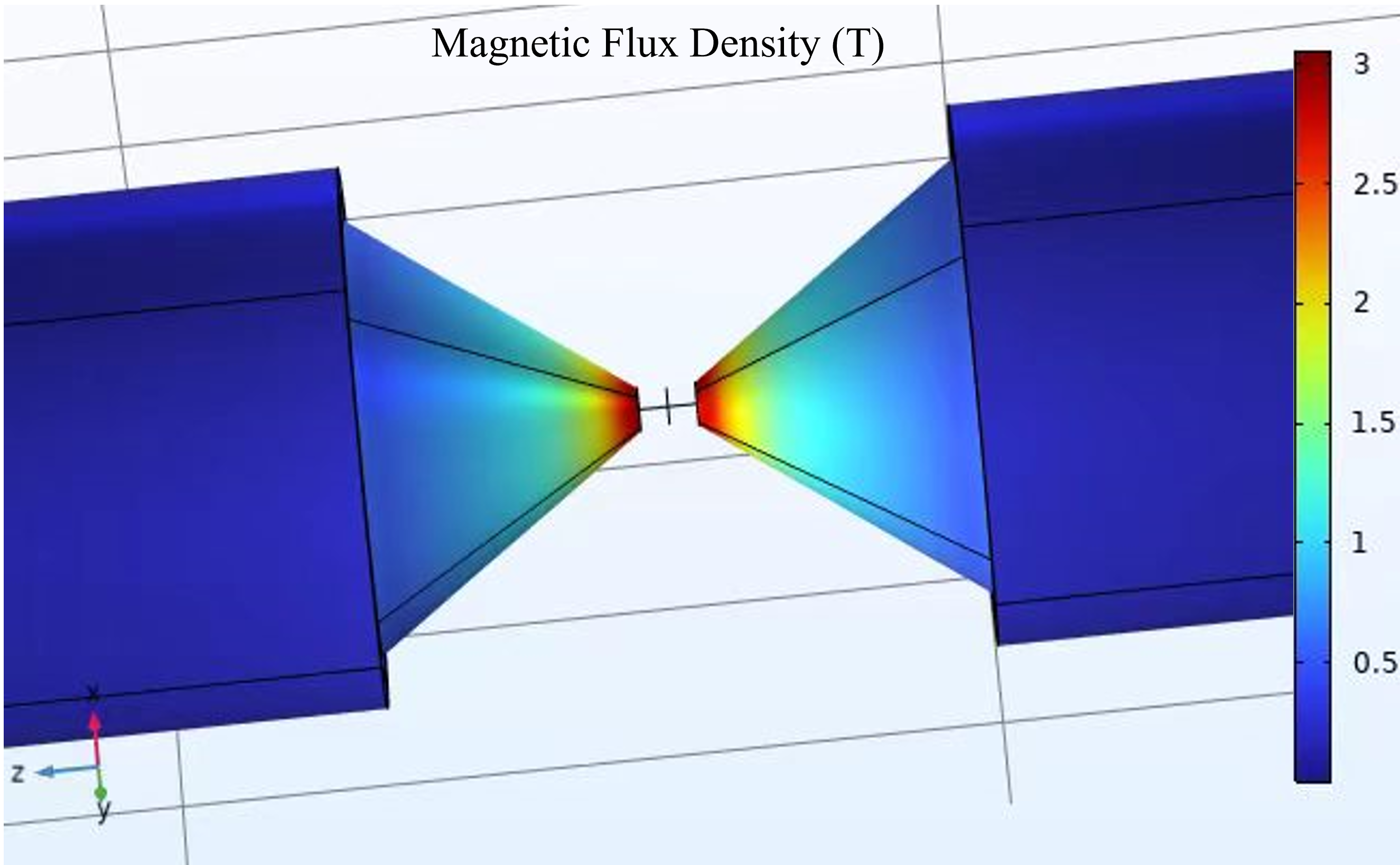}
\includegraphics[width=0.425\columnwidth]{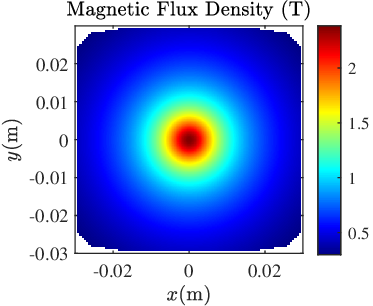}
    \caption{Left panel: the simulated magnetic field in three dimensions, where it is clear that the spearhead end does increase the magnetic flux density. Right panel: the magnetic flux density distribution in the $x-y$ plane which is the vertical plane between the electromagnets.}
    \label{simulation}
\end{figure}
We plot the geometry of our electromagnetic and simulate the produced magnetic field with COMSOL Multiphysics. For key parameters, we built the geometry of the core and the wire to be exactly equal to the real version, took the relative permeability of the silicon steel core to be 8000, an empirical value, and set the current to be 10A. As illustrated in the left panel of Fig. \ref{simulation} the magnetic field indeed concentrates to the spear-like ends of the electromagnet, making it much stronger than the rest of the electromagnet. 

Back in Fig. \ref{shadow}(a), after closer inspection, one can find that the shadow in the area between the electromagnet spearheads are lighter while the area above or below are darker. This is caused by the uneven distribution of magnetic field, as shown in Fig. \ref{simulation}. That is why in our quantitative measurement, we focus on the very centre area between the electromagnets, which is where our $B-I$ curve was measured as well.

\section{Mearsurement of the magnetic field strength of the electromagnet\label{B-I}}
\begin{figure}[H]
\includegraphics[width=1\linewidth]{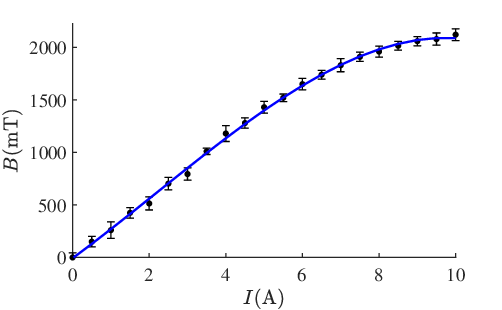}
\caption{The relation between the measured magnetic flux density $B$ and the current $I$ supplied to the electromagnet.}
    \label{BI}
\end{figure}

To quantitatively change the magnetic field at the centre, we measure the magnetic field strength of the electromagnet with the Tesla meter for different applied current. As shown in Fig. \ref{BI}, the obtained $I-B$ curve allows us to change current in the experiments and get the magnetic field strength corresponding to the current value.

\section{Theoretical analysis on shadow brightness\label{k-B}}In this section, we derive the shadow brightness factor $\kappa(B,\eta)$ as a function of the magnetic field strength $B$ and salt concentration $\eta$. The intensity of the shadow region can be expressed as 
\begin{equation}
K_{s}=K_{f}+K_{e},
\label{Ks}
\end{equation}
where $K_{f}$ is the intensity of the light emitted by the sodium lamp after passing through the flame and $K_{e}$ represents the light intensity of the background. Note that $K_{f}=K_{t}-K_{a}$ with $K_{t}$ being the brightness of the light emitted by the sodium lamp without absorption ($K_a=0$). Considering that the absorption intensity of sodium vapor for light of frequency $\omega$ is Gaussian-shaped \cite{gribl2021robust}

\begin{equation}
K_{a}(\omega)=\sum_{i=1,2}K_{\omega_{i}}e^{-\frac{\left(\omega-\omega_{i}\right)^{2}}{\Delta\omega^{2}}},
\end{equation}
where $\omega_{1}$ and $\omega_{2}$ are two main energy levels frequencies of sodium relevant to this phenomenon, corresponding to sodium lamp emission wavelengths of 589.6 nm and 590.0 nm. $\Delta\omega$ is the level width, and $K_{\omega_{i}}$ is the corresponding absorption intensity at resonance without magnetic field. After applying a magnetic field, the originally resonating sodium atomic levels $\omega=\omega_{i}$
will split due to Zeeman effect, resulting in \cite{jackson1938hyperfine,foot2005atomic}

\begin{equation}
\omega_{i,j_{i}}=\omega_{i}+\gamma_{i,j_{i}}B.
\end{equation}
Here, $\gamma_{1,j_{1}}$ ($1\leq j_{1}\leq4$) is proportional to the Lande $g$ factor $g_{1,j_{1}}=\pm2/3,\pm4/3$, while $\gamma_{1,j_{2}}$
($1\leq j_{2}\leq6$) is proportional to $g_{1,j_{2}}=\pm1/3,\pm1,\pm5/3$.
In this case, we obtain 

\begin{equation}
K_{a}=k\eta\sum_{i=1,2}\sum_{j_{i}}\tilde{K}_{\omega_{i}}e^{-\frac{\gamma_{i,j_{i}}^{2}B^{2}}{\Delta\omega_{i,j_{i}}^{2}}} ,
\label{Ka}
\end{equation}
where we have assumed that the intensity $K_{\omega_{i}}$ is proportional to (with $k$ the factor) the light intensity emitted by the sodium lamp $\tilde{K}_{\omega_{i}}$ and the concentration of sodium vapor which is further assumed to be proportional to the salt concentration $\eta$. Another assumption here is that the concentration of sodium atomic vapor inside the flame is proportional to the concentration of sodium in the salt water we mixed. The parameters $\tilde{K}_{\omega_{i}},\Delta\omega_{i,j_{i}}$
need to be obtained through precise spectroscopic information on sodium's emission and absorption spectra. 

For further analytical discussion, we focus on the regime of $\lambda_{i,j_{i}}B^{2}\ll1$ ($\lambda_{i,j_{i}}\equiv\gamma_{i,j_{i}}^{2}/\Delta\omega_{i,j_{i}}^{2}$), such that Eq. \eqref{Ka} is approximated as

\begin{align}
K_{a} & \approx k\eta\sum_{i=1,2}\sum_{j_{i}}\tilde{K}_{\omega_{i}}\left(1-\lambda_{i,j_{i}}B^{2}\right),\\
 & =k\eta\left(\sum_{i,j_{i}}\tilde{K}_{\omega_{i}}\right)e^{\ln\left(1-\frac{\sum_{i,j_{i}}\tilde{K}_{\omega_{i}}\lambda_{i,j_{i}}}{\sum_{i,j_{i}}\tilde{K}_{\omega_{i}}}B^{2}\right)} \\
 & \approx k\eta\left(\sum_{i,j_{i}}\tilde{K}_{\omega_{i}}\right)e^{-\frac{\sum_{i,j_{i}}\tilde{K}_{\omega_{i}}\lambda_{i,j_{i}}}{\sum_{i,j_{i}}\tilde{K}_{\omega_{i}}}B^{2}}\\
 & \equiv K_{0}\eta e^{-\lambda B^{2}}\label{Ka1}
\end{align}
where $K_{0}$ and $\lambda$ the effective parameters. It follows from Eqs. \eqref{Ks} and \eqref{Ka1} that

\begin{equation}
K_{s}(B,\eta)=K_{t}+K_{e}-K_{0}\eta e^{-\lambda B^{2}},
\end{equation}
and then we define the brightness factor of the shadow as

\begin{equation}
\kappa(B,\eta)\equiv\frac{K_{s}(B,\eta)}{K_{s}(0,0)}=1-A\eta e^{-\lambda B^{2}},
\end{equation}
with $A\equiv K_{0}/(K_{t}+K_{e})$. The above equation quantitatively demonstrates the dependence of shadow brightness on magnetic field strength and salt concentration, leaving
only two parameters $\lambda$ and $A$ to be determined by fitting experimental data. In our control variable experiment, for a given $\eta$, the specific fitting function associated with changing $B$ is $\kappa=1-b_1e^{-\lambda B^{2}}$; while for a given $B$, the specific fitting function associated with changing concentration is $\kappa=1-b_2\eta$.

\bibliography{ref}

\begin{thebibliography}{10}%
\makeatletter
\providecommand \@ifxundefined [1]{%
 \@ifx{#1\undefined}
}%
\providecommand \@ifnum [1]{%
 \ifnum #1\expandafter \@firstoftwo
 \else \expandafter \@secondoftwo
 \fi
}%
\providecommand \@ifx [1]{%
 \ifx #1\expandafter \@firstoftwo
 \else \expandafter \@secondoftwo
 \fi
}%
\providecommand \natexlab [1]{#1}%
\providecommand \enquote  [1]{``#1''}%
\providecommand \bibnamefont  [1]{#1}%
\providecommand \bibfnamefont [1]{#1}%
\providecommand \citenamefont [1]{#1}%
\providecommand \href@noop [0]{\@secondoftwo}%
\providecommand \href [0]{\begingroup \@sanitize@url \@href}%
\providecommand \@href[1]{\@@startlink{#1}\@@href}%
\providecommand \@@href[1]{\endgroup#1\@@endlink}%
\providecommand \@sanitize@url [0]{\catcode `\\12\catcode `\$12\catcode
  `\&12\catcode `\#12\catcode `\^12\catcode `\_12\catcode `\%12\relax}%
\providecommand \@@startlink[1]{}%
\providecommand \@@endlink[0]{}%
\providecommand \url  [0]{\begingroup\@sanitize@url \@url }%
\providecommand \@url [1]{\endgroup\@href {#1}{\urlprefix }}%
\providecommand \urlprefix  [0]{URL }%
\providecommand \Eprint [0]{\href }%
\providecommand \doibase [0]{https://doi.org/}%
\providecommand \selectlanguage [0]{\@gobble}%
\providecommand \bibinfo  [0]{\@secondoftwo}%
\providecommand \bibfield  [0]{\@secondoftwo}%
\providecommand \translation [1]{[#1]}%
\providecommand \BibitemOpen [0]{}%
\providecommand \bibitemStop [0]{}%
\providecommand \bibitemNoStop [0]{.\EOS\space}%
\providecommand \EOS [0]{\spacefactor3000\relax}%
\providecommand \BibitemShut  [1]{\csname bibitem#1\endcsname}%
\let\auto@bib@innerbib\@empty
\bibitem [{\citenamefont {Bakker}(1946)}]{bakker1946fifty}%
  \BibitemOpen
  \bibfield  {author} {\bibinfo {author} {\bibfnamefont {C.}~\bibnamefont
  {Bakker}},\ }\bibfield  {title} {\bibinfo {title} {Fifty years zeeman
  effect},\ }\href
  {https://www.sciencedirect.com/science/article/abs/pii/S0031891446800818}
  {\bibfield  {journal} {\bibinfo  {journal} {Physica}\ }\textbf {\bibinfo
  {volume} {12}},\ \bibinfo {pages} {555} (\bibinfo {year} {1946})}\BibitemShut
  {NoStop}%
\bibitem [{\citenamefont {Srivastava}\ \emph {et~al.}(2015)\citenamefont
  {Srivastava}, \citenamefont {Sidler}, \citenamefont {Allain}, \citenamefont
  {Lembke}, \citenamefont {Kis},\ and\ \citenamefont
  {Imamo{\u{g}}lu}}]{srivastava2015valley}%
  \BibitemOpen
  \bibfield  {author} {\bibinfo {author} {\bibfnamefont {A.}~\bibnamefont
  {Srivastava}}, \bibinfo {author} {\bibfnamefont {M.}~\bibnamefont {Sidler}},
  \bibinfo {author} {\bibfnamefont {A.~V.}\ \bibnamefont {Allain}}, \bibinfo
  {author} {\bibfnamefont {D.~S.}\ \bibnamefont {Lembke}}, \bibinfo {author}
  {\bibfnamefont {A.}~\bibnamefont {Kis}},\ and\ \bibinfo {author}
  {\bibfnamefont {A.}~\bibnamefont {Imamo{\u{g}}lu}},\ }\bibfield  {title}
  {\bibinfo {title} {Valley zeeman effect in elementary optical excitations of
  monolayer wse 2},\ }\href {https://www.nature.com/articles/nphys3203}
  {\bibfield  {journal} {\bibinfo  {journal} {Nature Physics}\ }\textbf
  {\bibinfo {volume} {11}},\ \bibinfo {pages} {141} (\bibinfo {year}
  {2015})}\BibitemShut {NoStop}%
\bibitem [{\citenamefont {Sun}\ \emph {et~al.}(2020)\citenamefont {Sun},
  \citenamefont {Cao}, \citenamefont {Cui}, \citenamefont {Zhu}, \citenamefont
  {Ma}, \citenamefont {Wang}, \citenamefont {Zhuo}, \citenamefont {Cheng},
  \citenamefont {Wang}, \citenamefont {Wan} \emph {et~al.}}]{sun2020large}%
  \BibitemOpen
  \bibfield  {author} {\bibinfo {author} {\bibfnamefont {Z.}~\bibnamefont
  {Sun}}, \bibinfo {author} {\bibfnamefont {Z.}~\bibnamefont {Cao}}, \bibinfo
  {author} {\bibfnamefont {J.}~\bibnamefont {Cui}}, \bibinfo {author}
  {\bibfnamefont {C.}~\bibnamefont {Zhu}}, \bibinfo {author} {\bibfnamefont
  {D.}~\bibnamefont {Ma}}, \bibinfo {author} {\bibfnamefont {H.}~\bibnamefont
  {Wang}}, \bibinfo {author} {\bibfnamefont {W.}~\bibnamefont {Zhuo}}, \bibinfo
  {author} {\bibfnamefont {Z.}~\bibnamefont {Cheng}}, \bibinfo {author}
  {\bibfnamefont {Z.}~\bibnamefont {Wang}}, \bibinfo {author} {\bibfnamefont
  {X.}~\bibnamefont {Wan}}, \emph {et~al.},\ }\bibfield  {title} {\bibinfo
  {title} {Large zeeman splitting induced anomalous hall effect in zrte5},\
  }\href {https://www.nature.com/articles/s41535-020-0239-z} {\bibfield
  {journal} {\bibinfo  {journal} {npj Quantum Materials}\ }\textbf {\bibinfo
  {volume} {5}},\ \bibinfo {pages} {36} (\bibinfo {year} {2020})}\BibitemShut
  {NoStop}%
\bibitem [{\citenamefont {Crutcher}\ and\ \citenamefont
  {Kemball}(2019)}]{crutcher2019review}%
  \BibitemOpen
  \bibfield  {author} {\bibinfo {author} {\bibfnamefont {R.~M.}\ \bibnamefont
  {Crutcher}}\ and\ \bibinfo {author} {\bibfnamefont {A.~J.}\ \bibnamefont
  {Kemball}},\ }\bibfield  {title} {\bibinfo {title} {Review of zeeman effect
  observations of regions of star formation},\ }\href
  {https://www.frontiersin.org/journals/astronomy-and-space-sciences/articles/10.3389/fspas.2019.00066/full}
  {\bibfield  {journal} {\bibinfo  {journal} {Frontiers in Astronomy and Space
  Sciences}\ }\textbf {\bibinfo {volume} {6}},\ \bibinfo {pages} {66} (\bibinfo
  {year} {2019})}\BibitemShut {NoStop}%
\bibitem [{\citenamefont {Kochukhov}(2021)}]{kochukhov2021magnetic}%
  \BibitemOpen
  \bibfield  {author} {\bibinfo {author} {\bibfnamefont {O.}~\bibnamefont
  {Kochukhov}},\ }\bibfield  {title} {\bibinfo {title} {Magnetic fields of m
  dwarfs},\ }\href
  {https://link.springer.com/article/10.1007/s00159-020-00130-3} {\bibfield
  {journal} {\bibinfo  {journal} {The Astronomy and Astrophysics Review}\
  }\textbf {\bibinfo {volume} {29}},\ \bibinfo {pages} {1} (\bibinfo {year}
  {2021})}\BibitemShut {NoStop}%
\bibitem [{\citenamefont {Taylor}\ \emph {et~al.}(2017)\citenamefont {Taylor},
  \citenamefont {Hyde},\ and\ \citenamefont {Batishchev}}]{taylor2017zeeman}%
  \BibitemOpen
  \bibfield  {author} {\bibinfo {author} {\bibfnamefont {A.~S.}\ \bibnamefont
  {Taylor}}, \bibinfo {author} {\bibfnamefont {A.~R.}\ \bibnamefont {Hyde}},\
  and\ \bibinfo {author} {\bibfnamefont {O.~V.}\ \bibnamefont {Batishchev}},\
  }\bibfield  {title} {\bibinfo {title} {Zeeman effect experiment with
  high-resolution spectroscopy for advanced physics laboratory},\ }\href
  {https://pubs.aip.org/aapt/ajp/article-abstract/85/8/565/1057974/Zeeman-effect-experiment-with-high-resolution?redirectedFrom=fulltext}
  {\bibfield  {journal} {\bibinfo  {journal} {American Journal of Physics}\
  }\textbf {\bibinfo {volume} {85}},\ \bibinfo {pages} {565} (\bibinfo {year}
  {2017})}\BibitemShut {NoStop}%
\bibitem [{\citenamefont {dos Santos}\ \emph {et~al.}(1996)\citenamefont {dos
  Santos}, \citenamefont {Porcher}, \citenamefont {Krupa},\ and\ \citenamefont
  {Gesland}}]{dos1996absorption}%
  \BibitemOpen
  \bibfield  {author} {\bibinfo {author} {\bibfnamefont {M.~C.}\ \bibnamefont
  {dos Santos}}, \bibinfo {author} {\bibfnamefont {P.}~\bibnamefont {Porcher}},
  \bibinfo {author} {\bibfnamefont {J.}~\bibnamefont {Krupa}},\ and\ \bibinfo
  {author} {\bibfnamefont {J.}~\bibnamefont {Gesland}},\ }\bibfield  {title}
  {\bibinfo {title} {Absorption and zeeman effect in-doped: measurements and
  simulation},\ }\href
  {https://iopscience.iop.org/article/10.1088/0953-8984/8/25/019/meta}
  {\bibfield  {journal} {\bibinfo  {journal} {Journal of Physics: Condensed
  Matter}\ }\textbf {\bibinfo {volume} {8}},\ \bibinfo {pages} {4643} (\bibinfo
  {year} {1996})}\BibitemShut {NoStop}%
\bibitem [{\citenamefont {Jackson}\ and\ \citenamefont
  {Kuhn}(1938)}]{jackson1938hyperfine}%
  \BibitemOpen
  \bibfield  {author} {\bibinfo {author} {\bibfnamefont {D.~A.}\ \bibnamefont
  {Jackson}}\ and\ \bibinfo {author} {\bibfnamefont {H.}~\bibnamefont {Kuhn}},\
  }\bibfield  {title} {\bibinfo {title} {The hyperfine structure of the zeeman
  components of the resonance lines of sodium},\ }\href
  {https://royalsocietypublishing.org/doi/abs/10.1098/rspa.1938.0127}
  {\bibfield  {journal} {\bibinfo  {journal} {Proceedings of the Royal Society
  of London. Series A. Mathematical and Physical Sciences}\ }\textbf {\bibinfo
  {volume} {167}},\ \bibinfo {pages} {205} (\bibinfo {year}
  {1938})}\BibitemShut {NoStop}%
\bibitem [{\citenamefont {Foot}(2005)}]{foot2005atomic}%
  \BibitemOpen
  \bibfield  {author} {\bibinfo {author} {\bibfnamefont {C.~J.}\ \bibnamefont
  {Foot}},\ }\href
  {https://global.oup.com/academic/product/atomic-physics-9780198506966?cc=ro&lang=en&}
  {\emph {\bibinfo {title} {Atomic physics}}},\ Vol.~\bibinfo {volume} {7}\
  (\bibinfo  {publisher} {Oxford university press},\ \bibinfo {year}
  {2005})\BibitemShut {NoStop}%
\bibitem [{\citenamefont {Gribl}\ and\ \citenamefont
  {Petrinovi{\'c}}(2021)}]{gribl2021robust}%
  \BibitemOpen
  \bibfield  {author} {\bibinfo {author} {\bibfnamefont {A.}~\bibnamefont
  {Gribl}}\ and\ \bibinfo {author} {\bibfnamefont {D.}~\bibnamefont
  {Petrinovi{\'c}}},\ }\bibfield  {title} {\bibinfo {title} {A robust method
  for gaussian profile estimation in the case of overlapping objects},\ }\href
  {https://www.aanda.org/articles/aa/abs/2024/05/aa49044-23/aa49044-23.html}
  {\bibfield  {journal} {\bibinfo  {journal} {IEEE access}\ }\textbf {\bibinfo
  {volume} {9}},\ \bibinfo {pages} {21071} (\bibinfo {year}
  {2021})}\BibitemShut {NoStop}%
\end{thebibliography}%

\end{document}